\documentclass[amsmath,amssymb,aps,prd,showpacs,nofootinbib]{revtex4}

\usepackage{graphicx}
\begin{document}

 \title{Recoupling Mechanism for exotic mesons and  baryons
 }
 \author{ \firstname{Yu.A.}~\surname{Simonov}}
\email{simonov@itep.ru} \affiliation{NRC ``Kurchatov Institute'' -- ITEP, B. Cheremushkinskaya 25, Moscow, 117259, Russia}


\newcommand{\beq}{\begin{eqnarray}}
 \newcommand{\eeq}{\end{eqnarray}}
\newcommand{\be}{\begin{equation}}
 \newcommand{\ee}{\end{equation}}

 \def\la{\mathrel{\mathpalette\fun <}}
\def\ga{\mathrel{\mathpalette\fun >}}
\def\fun#1#2{\lower3.6pt\vbox{\baselineskip0pt\lineskip.9pt
\ialign{$\mathsurround=0pt#1\hfil ##\hfil$\crcr#2\crcr\sim\crcr}}}
\newcommand{\veX}{\mbox{\boldmath${\rm X}$}}
\newcommand{{\SD}}{\rm SD}
\newcommand{\pp}{\prime\prime}
\newcommand{{\Mc}}{\mathcal{M}}
\newcommand{\veY}{\mbox{\boldmath${\rm Y}$}}
\newcommand{\vex}{\mbox{\boldmath${\rm x}$}}
\newcommand{\vey}{\mbox{\boldmath${\rm y}$}}
\newcommand{\ver}{\mbox{\boldmath${\rm r}$}}
\newcommand{\vesig}{\mbox{\boldmath${\rm \sigma}$}}
\newcommand{\vedelta}{\mbox{\boldmath${\rm \delta}$}}
\newcommand{\veP}{\mbox{\boldmath${\rm P}$}}
\newcommand{\vep}{\mbox{\boldmath${\rm p}$}}
\newcommand{\veq}{\mbox{\boldmath${\rm q}$}}
\newcommand{\veQ}{\mbox{\boldmath${\rm Q}$}}
\newcommand{\vez}{\mbox{\boldmath${\rm z}$}}
\newcommand{\veS}{\mbox{\boldmath${\rm S}$}}
\newcommand{\veL}{\mbox{\boldmath${\rm L}$}}
 \newcommand{\veA}{\mbox{\boldmath${\rm A}$}}
\newcommand{\veR}{\mbox{\boldmath${\rm R}$}}
\newcommand{\ves}{\mbox{\boldmath${\rm s}$}}
\newcommand{\vek}{\mbox{\boldmath${\rm k}$}}
\newcommand{\ven}{\mbox{\boldmath${\rm n}$}}
\newcommand{\veu}{\mbox{\boldmath${\rm u}$}}
\newcommand{\vev}{\mbox{\boldmath${\rm v}$}}
\newcommand{\veh}{\mbox{\boldmath${\rm h}$}}
\newcommand{\vew}{\mbox{\boldmath${\rm w}$}}
\newcommand{\verho}{\mbox{\boldmath${\rm \rho}$}}
\newcommand{\vexi}{\mbox{\boldmath${\rm \xi}$}}
\newcommand{\veta}{\mbox{\boldmath${\rm \eta}$}}
\newcommand{\veZ}{\mbox{\boldmath${\rm Z}$}}
\newcommand{\veB}{\mbox{\boldmath${\rm B}$}}
\newcommand{\veH}{\mbox{\boldmath${\rm H}$}}
\newcommand{\veE}{\mbox{\boldmath${\rm E}$}}
\newcommand{\veJ}{\mbox{\boldmath${\rm J}$}}
\newcommand{\veal}{\mbox{\boldmath${\rm \alpha}$}}
\newcommand{\vepi}{\mbox{\boldmath${\rm \pi}$}}
\newcommand{\vemu}{\mbox{\boldmath${\rm \mu}$}}
\newcommand{\vegam}{\mbox{\boldmath${\rm \gamma}$}}
\newcommand{\vepar}{\mbox{\boldmath${\rm \partial}$}}
\newcommand{\llan}{\langle\langle}
\newcommand{\rran}{\rangle\rangle}
\newcommand{\lan}{\langle}
\newcommand{\ran}{\rangle}
\newcommand{\oIm}{\operatorname{Im}}
\newcommand{\oRe}{\operatorname{Re}}
\newcommand{\vepsi}{\mbox{\boldmath${\rm \psi}$}}

 \Large

\begin{abstract}
The infinite chain of transitions of one pair of mesons (channel I) into another pair of mesons (channel II) can produce bound states and resonances in both channels even if no interactions inside channels exist. These resonances
which can occur also in meson-baryon channels are called channel-coupling (CC) resonances. A new mechanism of CC resonances is proposed where transitions occur due to a rearrangement of confining strings inside each channel --
the recoupling mechanism. The amplitude of this recoupling mechanism is expressed via overlap integrals of the wave
functions of participating mesons (baryons). The explicit calculation with the known wave functions yields the peak
at $E= 4.12$ GeV for the transitions $J/\psi + \phi  \leftrightarrow  D^*_s + \bar D^*_s $, which can be associated
with $\chi_{c1}(4140)$, and a narrow peak at $3.98$ GeV with the width $10$ MeV for the transitions
$D^{-}_s + D^*_0 \leftrightarrow J/\psi + K^{*-},$ which can be associated with th recently discovered $Z_{cs}(3985)$.

\end{abstract}

\maketitle

\section{Introduction}
The modern situation with the spectra of quarkonia and baryons  requires the dynamical
explanation of numerous extra states, which are not  present in the one-channel
spectra of a given  meson \cite{1}. A similar situation occurs in the excited baryon spectra \cite{1*}.  To be more precise,  in
  the case   of heavy quarkonia, i.e. states, which contain
$c\bar c$ and $b \bar b$ pairs,    the experimental data contain
a number of charged $Z_c, Z_b$ and neutral $Y_c, Y_b$ states, which cannot be
explained by the  dynamics of $c\bar c$, or $b\bar b$ pairs alone, see \cite{2}
for review.

There are theoretical  suggestions of different  mechanisms
\cite{3,4,5,6,7,8,9,10,
11,12,13,14,15,16,17,18,19,19*,19**,19***}, which should be taken into
account. E.g. poles (resonances) in the meson-meson  channels
can occur due to strong interaction in these systems, and appear as additional
poles in  the $S$ matrix  \cite{4,7,16,17} -- the molecular-type approach.

 A similar in  the choice of the driving channels $(Q\bar q\bar Q q)$, but
 different in the dynamics, is the approach of the  tetraquark  model \cite{ 5,9,11,12,15,18,19***}, see \cite{*23,**23, ***23} for reviews. A more general approach  contains features of both molecular and tetraquark models -- the hybridized tetraquark model \cite{*20}.
 One of the basic features of these models (as well as of the hadron-hadron interaction in general) is the determination of the dynamics in the system of two white objects. In the case of a deep tetraquark state one can
disregard the two-body white asymptotics with some accuracy,however in the molecular case and in general one needs
a classification of possible dynamical exchanges $V_{hh}$ between two white hadrons, taking into account the full gauge invariance and the confinement via general Wilson loop representation, which is now standard. It is clear that one-gluon exchange
and its spin versions (spin-spin,spin-orbit,tensor) are not present in $V_{hh}$, while two-gluon exchange
is the glueball exchange. The ``exchange of confining interaction'' is not a local potential but the transition of two    white objects into a common white one with $q\bar q$ annihilations before and after, e.g. $\pi + \pi \to \rho \to \pi + \pi$. In a similar way one obtains (actually nonlocal) $t-$channel and $u-$channel hadron exchanges.
A special role is played by the Coulomb (not color Coulomb)  interaction, which always exists locally in addition
to listed above interactions. It is important that all types of interaction between white objects (except for Coulomb) are short ranged for massive exchanged hadrons and do not produce narrow singularities by themselves.
Our purpose in this paper is to find the dynamical origin of relatively narrow peaks nearby threshold which occur
due to transformation of one white pair into another arbitrary number of times and to find what is the transformation itself. As we shall see, we shall meet a completely different type of transformation -- we shall call
it the recoupling and explain this mechanism in some detail.
 As a result we shall have here a new theoretical treatment of  hadron-hadron interaction, suggesting  a simple and   quite general mechanism for  exotic peaks in mesons and  baryons -- the  recoupling mechanism.

In contrast to the approaches, where white-white interaction in  the one-channel system, (e.g. in meson-meson)
is generating resonances, we propose the dynamical picture, where the summed up  transitions
  from one  channel  to another   (without interaction inside channels) can be strong enough to produce resonances  nearby
thresholds. The specific feature of this interaction is that it depends
strongly on the  wave functions of both channels, entering in the overlap integral of the transition  matrix
element, which measures the amplitude of the transition between the initial $q\bar Q+ \bar q Q$ and  final $q\bar q+Q\bar Q$ states.

 In what follows we are  exploiting the
 channel-coupling (CC) interaction in the form  of the energy-dependent  recoupling   Green's functions as a possible origin
 of extra states - the recoupling mechanism.

 Indeed, more than 30 years ago, the present author participated in the
 systematic study of CC effects in the  spectra of hadrons, nuclei and atoms
 \cite{20}. It was found there,
  that the CC interaction   defined  by the Transition Matrix Element (TME) is able to produce
 resonances (poles) of its own, if TME is strong enough, i.e. if the corresponding TME
 satisfies certain conditions, similar to  that for one-channel potential.

We show below, that at the basis of this recoupling process lies a simple picture of the string recoupling  between the same systems of quarks and antiquarks, which does not need neither energy nor  additional interaction, and is simply a kind of  topological transformation of two confining strings with fixed ends into another pair of strings -- the string recoupling.

\begin{figure}[!htb]
\begin{center}
\includegraphics[angle=0,width=8 cm]{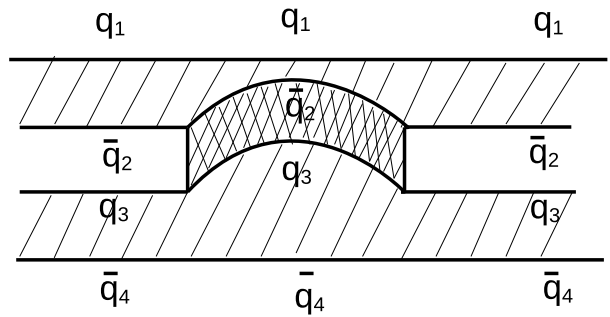}
\caption{ The transition of the mesons $(q_1\bar q_2) + (q_3\bar q_4) \leftrightarrow (q_1\bar q_4) + (\bar q_2 q_3)$
  via recoupling of the confining strings.
 } \vspace{1cm}
 \end{center}

\end{figure}

 One can see in Fig. 1 the confining regions (the crossed areas) for the bound states of quark-antiquark (mesons)
 $q_1 \bar q_2$ and $q_3 \bar q_4$ in the l.h.s. of the Fig. 1, which is transformed in the middle part of Fig. 1
 into the confining region between $q_1 \bar q_4$ on the plane of the figure, and ``the confining bridge" --
 the double-crossed area between $\bar q_2 q_3$. The r.h.s. of the Fig. 1 is the same as the l.h.s.
 As the result the transition
 is $(q_1\bar q_2) + (q_3\bar q_4) \leftrightarrow (q_1\bar q_4) + (\bar q_2 q_3)$.
 It is interesting to understand what kind of vertices are responsible for this transition, and to this end we
 demonstrate in Fig. 2 below the possible construction of the ``confining bridge" in the Fig. 1 by cutting the confining
 film and turning up the middle piece.

\begin{figure}[!htb]
\begin{center}
\includegraphics[angle=0,width=8 cm]{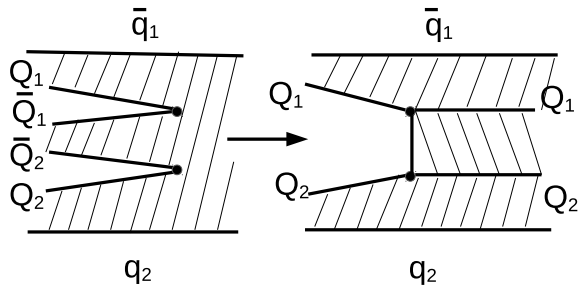}
\caption{The mechanism of string recoupling via double string-breaking process at the points shown by thick dots. } \vspace{1cm}
 \end{center}

\end{figure}

 We show in this way  that topologically this process is  equivalent to the double string breaking, and numerically is defined by the everlap integral of participating hadron wave functions. This mechanism is  quite general and can work for meson-meson, meson-baryon, baryon-baryon states. In particular it  can work  for some of $X,Y$ and $Z$ states of heavy  quarkonia,  like $Z_c(3900)$ and $Z_c(4020)$.

 It is a purpose of the present paper to exploit this formalism for the case of
 extra states  in meson-meson  or  meson-baryon spectra and define possible
   resonances  and thresholds, and  further on to apply this formalism to the case of pentaquark states like $P_c (4312), P_c(4440), P_c(4457)$.

 Our main procedure will be the  calculation of TME using realistic wave
 functions of $c\bar c$, $b \bar b$,  $c\bar u$ systems, as well as  approximate for baryon systems. Using those we calculate the  resulting Green's
 functions and resonance  positions  and compare them with experiment. The plan
 of the paper is as follows.

In the next section we introduce the reader to the method by solving a simplified
 two-channel problem with a  separable potential.   Section 3 is devoted to the explicit formulation of the  recoupling mechanism, section 4  contains  application  to the   meson-meson  channel, and section 5  considers the meson-baryon   case. Section 6, 7 are devoted to the analysis of the physical structure and numerical results and discussion, while section 7 contains
  conclusions and an outlook.

 \section{The simplest case: only separable CC interaction}

 Suppose we have two channels 1 and 2 with  thresholds $E_1$ and $E_2$ and  the
 CC interaction is  separable
 \be V_{12} (p_1, p_2) =- \lambda v_1 (p_1) v_2 (p_2) = V_{21}.\label{1}\ee
The Schroedinger-like (possibly relativistic) equations are \be (T_1
-E)\varphi_1 + V_{12} \varphi_2 =0, ~~(T_2 -E)\varphi_2 + V_{21} \varphi_1
=0\label{2}\ee
 and can be reduced to the equation
 \be (T_1
-E)\varphi_1 + V_{121} (E)\varphi_1 =0, \label{3}\ee where $V_{121}$ is \be
V_{121} (p_1, p_1'E) =- \lambda^2 v_1 (p_1) v_1 (p'_1) \int \frac{v_2^2 (k)
d^3 k/(2\pi)^3}{T_2(k) + E_2 - E}.\label{4}\ee

Solving (\ref{3}) one obtains the equation for the eigenvalue $E$ \be 1=
\lambda^2\int \frac{v_1^2 (p) d^3 p/(2\pi)^3}{T_1(p) + E_1 - E}\int \frac{v_2^2
(k) d^3 k/(2\pi)^3}{T_2(k) + E_2 - E},\label{5}\ee
  or  \be 1= \lambda^2 I_1 (E) I_2(E).\label{6}\ee
  For $E_1  < E_2$ one can put $E=E_1$ and get a condition for the existence of
  a bound state in our two-channel system \cite{20}.
  \be \lambda^2 I_1 (E ) I_2(E ) \geq 1.\label{7}\ee

  One of the intriguing  points now is how the bound state poles, or more
  generally, any poles appear when the interaction strength $\lambda$ is large
  enough. To this end we make a simplifying assumption about the form of $v_i
  (k)$ and write
  \be  a)~~ v_i^2 (k)= \frac{1}{k^2+ \nu^2_i} ~~~b)~~ v_i^2 (k)= \exp\left(-\frac{ k^2}{4\beta^2_i}\right)   \label{8}\ee
  where $\nu_1, \nu_2$ and $\beta_1,\beta_2$ are some constants. Assuming also the nonrelativistic
  kinematics $T_i =\frac{k_1^2}{2\mu_i}$ one obtains in the case  a)
  \be I_i (E ) = \frac{\mu_i}{2\pi (\nu_i - i \sqrt{2\mu_i \Delta_i})}, ~~
  \Delta_i = E-E_i,\label{9}\ee
  and we have taken square root   (\ref{9}) on the physical Riemann sheet, $
  E=E+i\delta$. Hence the equation (\ref{5}) for the poles (energy eigenvalues)
  is
  \be (\nu_1 - i \sqrt{ 2\mu_1 (E-E_1)}) (\nu_2 - i \sqrt{2 \mu_2 (E-E_2)}) =
  C=\frac{\mu_1\mu_2\lambda^2}{4\pi^2}.\label{10}\ee
  We shall be mostly interested in the poles around the threshold $E_2$ and
  therefore in the first approximation we replace the first factor on the
  l.h.s. of (\ref{10}) by a  constant, assuming, that $E_2-E_1$ has a large
  positive value, hence one can write for $k_2 = \sqrt{\mu_2 (E-E_2)}$ using
  (\ref{10})\be k_2 \cong - i\nu_2 + i \lambda^{'2}, ~~ \lambda^{'2} =
  \frac{\mu_1\mu_2\lambda^2}{4\pi^2 (\nu_1 - i \sqrt{2\mu_1
  (E_2-E_1)})}.\label{11}\ee

  From (\ref{1}) one can see that the pole is originally (at $\lambda'=0$) on
  the second $E_2$ sheet,  $k_2=-i\nu_2$ and remains on  the second $E_2$
  sheet with increasing $\lambda'$. Note, however, that since originally we
  have been on the $E_1$ first sheet, then  $\oIm\lambda^{'2}>0$, and therefore
  $\oRe k_2<0$, implying that the pole can be  of the  Breit-Winger type for $\oRe \lambda^{'2}>\nu_2$ .

  As will be shown below in section 4, resonance production cross sections are
  proportional to the function
  \be
  \frac{d\sigma^* (E)}{dE} = \left| \frac{1}{1- \lambda^2 I_1 (E-E_1) I_2 (E-E_2)}
  \right|^2 k_2(E).\label{12}\ee

  We can  generalize this separable form to the  relativistic  case, when two hadrons with masses $m_3,m_4$, so that  the  denominators  in (\ref{5}) look as follows:
   $$ T_1(p) + E_1 -E \to \sqrt{p^2 + m^2_1} + \sqrt{p^2+m^2_2} -E,$$
   \be  T_2(p) + E_2 -E \to \sqrt{k^2 + m^2_3} + \sqrt{k^2+m^2_4} -E.\label{12*}\ee

   Here we have two thresholds $m_1+m_2$ and $m_3+m_4$,  and we shall assume that $m_1+m_2< m_3+m_4$.

   Making the  replacement (\ref{12*}) in $I_1(E), I_2(E)$ one can calculate these functions and find the behaviour of the approximate cross section in (\ref{12}).

\section{Equations for two channel amplitudes in the  recoupling formalism}

In this section we discuss
 the Green's function
  of the  system of two white (noninteracting) hadrons $h_1,h_2$, which can transform into another system  of white hadrons $H_3,H_4$ and this transformation   can occur infinite number of times $h_1h_2 \to H_3H_4\to h_1h_2 \to H_3H_4\to ...$

  Denoting the transition amplitude $V(h_1h_2 \to H_3H_4)= V^+ (H_3H_4\to h_1h_2)$, and the corresponding Green's functions as $G_h, G_H$,  we obtain the   total  Green's function $G_{\alpha\beta}$, e.g. $G_{hh}$

  $$ G_{hh} = G_h + G_hV_{hH} G_H V_{Hh} G_h+G_hV_{hH} G_H V_{Hh} G_h V_{hH} G_H V_{Hh} G_h+ ... =$$\be =G_h \frac{1}{1- V_{hH} G_HV_{Hh}G_h}; \label{1a}\ee
  as a result one obtains the equation,
 which defines all possible singularities of the physical amplitudes, including the resonance poles.
 \be 1=V_{hH} G_H V_{Hh} G_h.\label{2a}\ee

 Note, that the  described above method of the channel coupling was proposed before in the nonrelativistic form by the Cornell group \cite{20*}, and  exploited for the analytic calculation of the charmonium spectra, where the $h_1h_2 $ are  strongly interacting quarks $c\bar c$. The subsequent  development of this  method  in \cite{21,22,23} has allowed  to understand the  nature of the  $X(3872)$ \cite{22} and $Z_b$ states \cite{23}. For the light quarks this method requires the explicit knowledge of $q\bar q$ spectrum and wave functions, which are available in the QCD string approach \cite{23*,30*}.

 Recently the same approach, called the relativistic Cornell-type formalism successfully explained the spectrum of light scalars \cite{23***,23**}. In our present case we disregard the interaction of hadrons $h_1$ with $h_2$ and $H_3$ with $H_4$.

 Both Green's function $G_H, G_h$  describe propagation of two  noninteracting subsystems, but each of these hadrons can have its own nontrivial spectrum.

 In the simplest case, e.g. $h=\pi\bar \pi, H=K\bar K$, the Green's functions of noninteracting particles are well known, see e.g. \cite{23***,23**} for  the scalar $\pi\pi, K\bar K$ Green's functions  with the  fixed spatial  distance between $\pi\pi$ or $K\bar K$, needed to define the  the transition matrix  element.

 Since each of $h_i$ or $H_j$  is a  composite system consisting of $q\bar q$ or $qqq$  one must write the corresponding relativistic composite Green's function, using the path integral formalism, see \cite{30*} for a recent review.

 As it is seen from (\ref{12*}), one needs the explicit form of the  relativistic Green's function, consisting of two  quark-antiquark mesons $ h_1(q_1, \bar Q_1)$ and $ h_2(\bar q_2, Q_2)$ with  the zero total momentum $\veP =0$, so that the c.m. momentum of $q_1\bar Q_1$ is  $\vep_1$, while for $\bar q_2Q_2$ it is $-\vep_1$. As a result the wave function of the $h_1h_2$ system with $\veP=0$ and c.m. coordinates $\veR$ can be written as

 \be \Psi_{h_1h_2} (\veu-\vex; \vey-\vev)= e^{i\vep_1 \veR (\veu, \vex)-i\vep_1 \veR (\vey, \vev)} \psi_1 (\veu-\vex) \psi_2(\vey-\vev).\label{32}\ee
 At the same time the relativistic wave function of the hadrons $H_3,H_4$,  $h_H(q_1\bar q_2), H_4(\bar Q_1 Q_2)$ has the form

 \be \Psi_{h_3h_4} (\veu-\vev; \vev-\vey)= e^{i\vep_2 \veR (\veu, \vev)-i\vep_2 \veR (\vex, \vey)} \psi_3 (\veu-\vev) \psi_4(\vex-\vey).\label{33}\ee

 Here we have introduced the c.m. coordinates  $\veR$ of the hadrons, expressed via the average energies $\omega_i, \Omega_i$ of the quarks and antiquarks in the hadron \cite{30**}
 \be \veR (\veu, \vex) = \frac{\omega_1 \veu+\bar \Omega_1\vex}{\omega_1 +\bar \Omega_1}, ~~ \veR (\vey, \vev) = \frac{\bar\omega_2 \vev+  \Omega_2\vey}{\bar \omega_2 +  \Omega_2},\label{34}
 \ee
 \be \veR (\veu, \vev) = \frac{\omega_3 \veu+\bar \omega_4\vev}{\omega_3 +\bar \omega_4}, ~~ \veR (\vex, \vey) = \frac{\bar\Omega_3 \vex+  \Omega_4\vey}{ \bar \Omega_3 +  \Omega_4}.\label{35}
 \ee

 Here $\omega_i,\Omega_i$ are given in the Appendix 1 of this paper.

 Next we must calculate the overlap matrix element of $\Psi_{h_1h_2}$ and $\Psi_{h_3h_4}$
 \be V_{12|34} (\vep_1, \vep_2)= \int \bar y_{1234} d^3 (u-x) d^3 (y-v) d^3 (u-v) \Psi_{h_1h_2} \Psi^+_{h_3h_4}.\label{36}\ee
 Introducing the Fourier component of the wave functions e.g.
 $\psi_1 (\veu-\vex) = \int \tilde \psi_1 (\veq_1) e^{i\veq_1 (\veu-\vex)} \frac{d^3q_1}{(2\pi)^3}$, one obtains in the simple case when $q_2=q_1, Q_2=Q_1$

  $$V_{12|34} (\vep_1, \vep_2)= \int \frac{d^3\veq_1}{(2\pi)^3}\bar y_{1234} \tilde \psi_1(\veq_1)\tilde \psi_2(\veq_1+\vep_2) \tilde \psi_3\left( - \veq_1 - \frac{\vep_2}{2} - \vep_1 \frac{\omega_1}{\omega_1+ \Omega_1}\right)\times$$\be\times\tilde\psi_4\left(   \veq_1 - \frac{\vep_2}{2} - \vep_1 \frac{\Omega_1}{\omega_1+ \Omega_1}\right).\label{37}\ee

In (\ref{36}), (\ref{37}) we introduced the numerical  recoupling coefficient $\bar y_{1234}$,  which is  discussed in  Appendix 2.

The transition element (\ref{36}) with  the factor $\bar y_{1234}$, responsible for the recoupling of hadrons, shown in Fig. 1, has a simple structure. Indeed, as one can see in Fig. 1, the creation of two string configurations in the intermediate confining  strings position and back into the original configuration.
One may wonder what is the explicit mechanism of this recoupling, and what are the vertices denoted by thick points in the Fig. 2. To this end we note, that we have two strings on  the r.h.s. of Fig. 2:
 string from $ Q_1$ to $\bar Q_2$ and another from $ \bar q_1$ to $ q_2$; this position  results from the double string decay  (the l.h.s. of  Fig. 2)  with the subsequent  rotation of the string between $ Q_1$ and $\bar Q_2$ to the  right, where this  string is  at some  distance above the string between $\bar q_1,q_2$.  One can associate the quantity $M(x,y)$ with this process and we must add this factor to $V_{12|34}$. Writing $M(x,y) = \sigma|\vex-\vey|$ in analogy with the one-point string decay described by the effective Lagrangian \cite{33***} for the string decay,

\be {\cal{L}}_{sd} = \int d^4x\bar \psi(x) M(x) \psi(x)\label{38}\ee
and replacing it  with the numerical value $M_{\omega}$, similarly to \cite{21,22,23}, (see Appendix 2 for  details)  one  can write $y_{1234}=M_\omega \chi_{1234}$ in (\ref{37}), with $\chi_{1234}$ describing the spin-isospin  recoupling. Finally  one  obtains the expression for the whole combination
\be N(E) = G_{h_1h_2} V_{12/34} G_{h_3h_4} V_{34/12}\label{39}\ee
\be N(E)= \int \frac{d^3\vep_1}{(2\pi)^3}   \frac{d^3\vep_2}{(2\pi)^3}\frac{V_{12|34} (\vep_1, \vep_2)V_{34|12 } (\vep_1, \vep_2)}{(E_1(\vep_1)+E_2(\vep_1)-E) (E_3(\vep_2)+E_4(\vep_2)-E)}.\label{40}\ee

 The resulting singularities (square root threshold singularities and possible poles from the equation $N(E)=1)$ can be found in the integral (\ref{40}).

 One can see, that the structure of the expression (\ref{40}) is the same as in Eq.(\ref{5}),  provided $V_{12|34} $ factorizes in factors $v_1( p_1) v_2(p_2)$, and consequently one expects the same behaviour of the cross sections as in (\ref{12}).

 At this point it is useful to introduce the approximate form of the wave functions in (\ref{37}), which is  discussed in \cite{23}. Here we only  give the simplest form of the Gaussian wave functions for the ground states of light, heavy-light and heavy  quarkonia. One can write
 \be \tilde \psi_i (q) = c_i \exp \left( - \frac{q^2}{2\beta^2_i}\right), ~~ c^2_i = \frac{8 \pi^{3/2}}{\beta_i^3};~~ \int \tilde \psi^2_i (q) \frac{ d^3q}{(2\pi)^3} =1\label{41}\ee
 where $\beta_i$ was found in \cite{21,22,23},  see Appendix 3, e.g. for ground states of bottomonium  $\beta =1.27$ GeV, for  charmonium $\beta=0.7$ GeV and for $D,B$ mesons $\beta =0.48, 0.49$ GeV.

 Inserting $\tilde \psi_i(q)$ in (\ref{41}) into (\ref{37}) and integrating  over $d^3\veq_1$ one obtains
 \be V_{12|34} (\vep_1\vep_2) = \bar y_{1234} \left(\prod^4_{i=1}  c_i\right)\frac {\exp (-AP^2_2 - Bp^2_1 -C\vep_1\vep_2)}{(2\pi)^3 8\pi^{3/2} a^{3/2}},\label{42}\ee
 where $a,A,B,C$ are
 \be a=\frac12 \sum^4_{i=1} \frac{1}{\beta^2_i}; ~~A= \frac12 \left( \frac{1}{\beta^2_2} + \frac{1}{4 \beta^2_3} + \frac{1}{4\beta^2_4} \right) - \frac{1}{4a} \left( \frac{1}{\beta^2_2} + \frac{1}{2 \beta^2_3} + \frac{1}{2\beta^2_4} \right)^2\label{43}\ee

 \be B=\frac{1}{2\beta^2_3} \left( \frac{\omega_1}{\omega_1+\Omega_1}\right)^2+
\frac{1}{2\beta^2_4} \left( \frac{\Omega_1}{\omega_1+\Omega_1}\right)^2-
\frac{1}{4a} \left(\frac{1}{\beta^2_3} \frac{\omega_1}{\omega_1+\Omega_1}-
\frac{1}{ \beta^2_4 } \frac{\Omega_1}{\omega_1+\Omega_1}\right)^2\label{44}\ee

 $$  C=   \frac12\left(\frac{1}{ \beta^2_3}  \frac{\omega_1}{(\omega_1+\Omega_1)}-
\frac{1}{ \beta^2_4}   \frac{\Omega_1}{(\omega_1+\Omega_1)}\right) -\frac{1}{2a }
\left( \frac{1}{\beta^2_2} + \frac{1}{2 \beta^2_3} + \frac{1}{2\beta^2_4} \right)
 $$
\be
 \left( \frac{\omega_1}{\beta^2_3(\omega_1+\Omega_1}-
  \frac{\Omega_1}{\beta^2_4(\omega_1+\Omega_1}\right). \label{45}\ee

 The resulting $N(E)$ has the form
 \be N(E) = \frac{ M^2_\omega \bar \chi^2_{1234}}{a^3(\prod_i \beta_i)^3(\pi)^3 }
 \int \frac{ d^3p_1
 d^3p_2 \exp  ( -  2A p^2_2-2Bp^2_1-2C\vep_1\vep_2     )}{(E_1(p_1) + E_2(p_1) -E) ( E_3(p_2)+ E_4(p_2)- E)}\label{48c}\ee
 and the differential cross section  with the  final second channel is proportional to
 \be \frac{d\sigma}{dE} \sim \frac{p (E)}{|1-N(E)|^2}\label{30c}\ee
 where $p(E)\sim \sqrt{E^2-(m_3+m_4)^2}$.
 It is interesting, that for the fully symmetric case, when all $\beta_i$ are equal, and  $\omega_1=\Omega_1$, one obtains for the exponent in (\ref{42}) $\exp \left( - \frac{p^2_1 + p^2_2}{2\beta^2}\right)$, and $V_{12|34}=V_{34|12}$ and $N(E)$ are
 \be
 V_{12|34}^{\rm symm}(\vep_1,\vep_2)= \frac{2^{5/2} \sqrt{\pi}}{\beta^3} \bar y_{1234} \exp \left( -\frac{p^2_1+p^2_2}{4\beta^2}\right)\label{46}\ee

 \be N(E) = \frac{2 M^2_\omega \bar \chi^2_{1234}}{\pi \beta^6} \int \frac{p^2_1 dp_1
 p^2_2 dp_2 \exp \left( - \frac {p^2_1+p^2_2}{4\beta^2}\right)}{(E_1(p_1) + E_2(p_1) -E) ( E_3(p_2)+ E_4(p_2)-(E)}.\label{47}\ee

\section{Recoupling mechanism for the meson-meson  amplitudes}

The formalism introduced on the previous section can be directly applied to the  amplitudes, containing two meson-meson thresholds, $m_1+m_2\leftrightarrow m_3+m_4$ with the  singularities given by the  equation
\be 1- N(m_1,m_2,m_3,m_4; E) =0.\label{48a}\ee
As we discussed in section II, the conditions for the appearance of  visible singularities require that the threshold difference $\Delta M= m_3+m_4-m_1-m_2$  should be  comparable or smaller than average size $\lan \beta\ran$ of the hadron wave functions in momentum space, while the recoupling coefficient $\bar y^2_{1234}$ is of the order of unity, i.e. there should be no   angular momentum excitation    or spin  flip   process.

An additional requirement is the relatively small widths of  participating hadrons, otherwise all singularities would be  smoothed out.

One can choose several examples in this respect.
\begin{description}
\item[1)]
 The set of tranformations
\be J/\psi+\phi \leftrightarrow D^*_s + \bar D^*_s \to J/\psi+\phi\label{49a}\ee
with masses  $m_1=3097$ MeV, $m_2=1020$ MeV, $m_3=m_4=2112$ MeV, and the corresponding thresholds $m_1+m_2=4117$ MeV and $m_3+m_4 =4224$ Mev. One can see no spin flip in the  sequence $c_+\bar c_+ + s_+\bar s_- \to c_+\bar s_- +\bar c_+ s_+$ for (\ref{48a}), where lower indices denote spin projections, and therefore no damping of transition probability. One can expect, that the yield of the reaction (\ref{49a}) would have the  form similar to that of $\chi_{c1}(4140)$ with the mass $(4147$ MeV  with the width $\Gamma= (22 \pm 8 )$ MeV  \cite{37c}.

\item[2)] One of the best studied exotic resonances $Z_c(3900) $ \cite{37.4,37.5,37.6,37.7}   was found in the reaction $e^+e^-\to \pi^+\pi^- J/\psi\to \pi^\pm Z_c(3900)$. It can be associated with the recoupling process $D\bar D^* \leftrightarrow \pi J/\psi$, where the higher threshold  is $M_2=3874$ MeV, and the spin, charge and isospin  recombination agrees with this recoupling. One expects the peak above $M_2$ in agreement with experiment.

A similar situation can be  in the case of the $Z_c(4020) $ observed in the reaction $e^+e^- \to \pi\pi h_c $\cite{37.8}, which can be  associated with the recoupling $\pi h_c\to D^* \bar D^*$ with threshold $M_1=3665$ MeV and  $M_2=4020$  MeV. One can  one can  envisage the yield of the  reaction to be  described  by  the equation (\ref{30c}), with $p(E) \to p^3(E)$, since one need the $P$-wave in $D^*\bar D^*$ near threshold, as in $h_c$.

\end{description}
Note, that in general the recoupling can easily produce both  $Z_b,Z_c$  resonance peaks, when a charged particle (like $\rho$) is participating in the sequence of transformations.

\section{Recoupling mechanism for meson-baryon systems}

One can consider the transformation sequence for baryons of the form, e.g.
\be (qqq) + Q\bar Q) \leftrightarrow (qqQ) + (q\bar Q)\label{37b}\ee
and apply the same formalism as the used above for the meson-meson recoupling transformations.

In principle it implies the new degrees of freedom, associated with  the additional quark in ($qqq)$ as compared to the meson $(q\bar q)$. To simplify the matter, we start below with the assumption, that the diquark combination can be factorized out in the baryon $(qqq) \to q(qq)$ and does not change during the recoupling process, which can now be written as
\be q(qq) + (Q\bar Q)  \leftrightarrow Q(qq ) + (q\bar Q).\label{38b}\ee
In doing  so we neglect also the internal structure of the diquark $(qq)$ system, which stays unchanged during the recoupling process, so that only its total spin, spin projection and its relative motion with the quark $q$ or$Q$ in the baryon bound state is present in the matrix element (\ref{40}), while the norm of $(qq)$ is factored out. As a result one can use eqs. (\ref{40}), (\ref{42}), where we need the wave functions of the  relative motion of quark and diquark in the baryons $q(qq)$ and $Q(qq)$. Using our  notations $h_1(q_1\bar Q_1) + h_2( \bar q_2 Q_2)\leftrightarrow h_3 (q_1\bar q_2 ) +h_4 ( \bar Q_1Q_2)$ we are  replacing $\bar q_2$ by the diquark $(qq')$. The accuracy of this replacement was discussed in literature  \cite{35,36,37,38,39} and  the interactions are discussed and compared in \cite{40}.
In what  follows we need the approximate baryon wave functions as $\psi_2(\vey-\vev) e^{-i\vep_1\veR(\vey,\vev)}$ and $\psi_3(\veu-\vev) e^{-i\vep_2\veR(\veu,\vev)}$ in   (\ref{2a}),(\ref{33}), where $\vev$ denotes the center-of-mass of the diquark, and $\veR(\vey,\vev)$ is the c.m. of the quark-diquark combination, i.e. actually is the c.m. of the baryon $Q_2(qq)$. The same for the $\psi_3(\veu-\vev)$ and its c.m. $\veR(\veu,\vev)$. Using the  oscillator forms for $\psi_2,\psi_3$ one is actually exploiting the  description of the  only one part (factor) of the baryon wave function, which can be  associated with only one leaf of the three-leaf baryon configuration. As a result, one can approximate this part of the wave function with  the wave function of the heavy-light meson  for the $Q_2(qq)$ baryon $(Q_2(qq) \to Q_2\bar q)$ or  with the light baryon for the  $h_3(q_1\bar q_2 \to q_1(qq))$.

As a first example one can take the transitions $p+\phi \to \bar K^* + \Lambda$ with thresholds $M_1=1960$ MeV and $M_2=2005$ MeV, where the role of quarks $\bar Q_1, Q_2$ is played by $\bar s, s$ and one has a  transition $u(ud) + \bar ss\to \bar su+s(ud)$, where all $\beta$ parameters have a similar magnitude, and one can expect a peak nearby $M_2$.

As a concrete example one can take the case of the triad  $P_c(4312), P_c(4440)$ and $P_c(4457)$, found experimentally in \cite{41,42}, with a vast literature devoted to this phenomenon, called pentaquark, see a   review in PDG \cite{1*},  and for the latest pentaquark papers see \cite{44,45,46,47,48,49,50,51,52,
53,54,55,56,57,58,59,63,64}.
.

The most part of the literature considers pentaquarks as a result of molecular interaction between a white   baryon  and  a  white meson, which creates a bound state nearby the threshold of this system.  In what follows we shall exploit the recoupling mechanism and we shall show, that it can provide the  observed peaks without an  assumption of the white-white strong interaction.

We shall have in mind the recoupling transformations of the type
\be J/\psi + P \leftrightarrow (\Sigma, \Sigma^*) +(D,D^*)\label{39b}\ee
and we impose the  requirement of $s$-wave recoupling without spin flip processes and parity  conservation, which excludes $\Lambda^*_c(2595)$ with $(IJ^P)=(0, \frac12^-)$ and includes $\Lambda_c(2286) (0, \frac12^+)$, $\Sigma_c(2455)(1, \frac12^+), \Gamma \approx 2$ MeV, $\Sigma_c^*(2529)(1, \frac32^+), \Gamma \approx 15$ MeV, in addition to $D(1864), (\frac12, 0^+)$ and $D^*(2010), (\frac12, 1^-)$ with $\Gamma_D, \Gamma_{D^*}<1$ MeV.

As a result one obtains the thresholds $M_2=m_3+m_4$ in the Table I together with $P_c$.

 \begin{table}[!h]
\caption{Meson-baryon thresholds and the associated pentaquarks }
 \begin{center}
\label{tab.01}\begin{tabular}{|c|c|c|c|c|}\hline
thresholds (MeV)&4150&4319&4465&4384\\ \hline
pairs &$\Lambda_c\bar D$& $\Sigma_c\bar D$&$\Sigma_c\bar D^*$&$\Sigma_c^*\bar D$\\\hline

pentaquarks&&$P_c(4312)$& $P_c(4457)$&$P_c(4440)$\\
 width, MeV&&$\Gamma=9.8$&$\Gamma=6.4$&$\Gamma=20.6$\\\hline

\end{tabular}

\end{center}
\end{table}
 Following the Table, we can consider two types of reactions,
 \be \begin{array}{cc} I.& (c(ud)) (\Sigma_c, \Sigma^*_c)+(\bar c u) (\bar D, \bar D^*) \leftrightarrow (c\bar c)(J/\psi) + (u(ud)) (p)\end{array}\label{40b}\ee
 \be \begin{array}{cc} II.& (c(uu)) (  \Sigma^*_c)+ \bar c d  (\bar D, \bar D^*) \leftrightarrow (c\bar c) + (u  (ud )) (p)\end{array}.\label{41b}\ee
 To proceed one needs the values of $\beta_{i}, i=1,2,3,4$ and  $A,B,C$ and $a$ in (\ref{48c}). Using Appendix 1 one finds the values of $\omega, \Omega$ in (\ref{43}), (\ref{44})
 $\Omega_1=1509$ MeV, $\omega_1=507$ MeV. From Appendix 3 one finds the values of $\beta_i$:
 \be
 \beta_i (D,D^*)\cong  \beta_2(\Sigma) = 0.48 ~{\rm GeV},~~\beta_3(p) \approx 0.26 ~{\rm GeV},\label{42b}\ee
 and  $\beta_4(J/\psi) = 0.7 ~{\rm GeV}$. Note, that for $\Sigma$ and $p$ we have used the principle of replacement of light diquark by a light antiquark, $(ud)\to \bar q$. Therefore $\beta(\sigma) = \beta (c(ud) = \beta (c\bar u)=\beta(D)$.

 As a consequence one obtains the values given in (\ref{42b}). Using those we get  the numerical values for $a,A,B,C $.
 \be  a= 12.76 ~{\rm GeV}^{-2}, ~~ A=4.02~ {\rm GeV}^{-2}, B=0.94~ {\rm GeV}^{-2}, ~~ |C|<0.03~ {\rm GeV}^{-2}.\label{43b}\ee

 As a result one can neglect the $C\vep_1\vep_2$ in (\ref{48c}) and the integrals $d^3\vep_1,d^3\vep_2$ factorize.

 We turn now to the recoupling coefficients $M_\omega, \bar y_{1234}$.

 As it was shown in \cite{33***}, the effective parameter $M_\omega$ can be expressed via the wave  functions of objects, produced by the string breaking, in our case it is heavy-light mesons with the coefficient $\beta(D)\cong \beta(B) =0.48$ GeV, and from Eq. ({35}) of \cite{33***} one has
 \be M_{\omega}\cong  \frac{2\sigma}{\beta(D)}\cong 0.8~{\rm GeV}.\label{44b}\ee
 Finally, the coefficient $\bar \chi_{1234}$ for the transition into $(\Sigma_c\bar D)$ and $(\Sigma_c\bar D^*)$ can  be estimated as in the Appendix 2 to be  equal to 1.

\section{ Physical amplitudes and their singularities in the recoupling process }

As was discussed in {section 3}, (\ref{30c}), the differential cross section for the production of hadrons in channel 1 can be written as
\be \frac{d\sigma}{d E}= |F_1(E) f_{12}(E)|^2 p_1(E)
\label{6.1}\ee
where $F_1(E)$ is the production amplitude of channel 1 particles without final state $FS$ interaction and $f_{12}$ is the $FS$ interaction,which we take as an infinite sum of transitions from channel 1 to channel 2-the Cornell-type
mechanism \cite{20*,21,22}.

$f_{12}(E)$ can be written as
\be f(E) = \frac{1}{1 - N(E)}
\label{6.2}\ee
where $N(E)$ has the form
\be
 N(E)= \lambda I_1(E) I_2(E),
 \label{6.3} \ee
 where $I_i(E)$ has the form
 \be
 I_i(E)= \int{\frac{d^3 p_i}{(2\pi)^3}\frac{v_i^2(p_i)}{E'(p_i) + E"(p_i) - E}}.
 \label{6.4} \ee
Here $v_i$ is proportional to the product of wave functions in momentum space (see (ref {42})) and can be written in two forms:
 a/ as a Gaussian of $p$ and b/ as an inverse of $(p^2 + \nu^2)$.
To simplify matter we shall consider situation close to nonrelativistic for the energies in the denominator of (\ref{6.4}) and write
\be E'(p_1) + E"(p_1)= m_1 + m_2 + \frac{p_1^2}{2\mu_1}, E'(p_2) + E"(p_2)= m_3 + m_4 +\frac{p_2^2}{2\mu_2}.
\label{6.5} \ee
We have considered above in the paper $v_i^2(p)$ as a Gaussian $\exp(-b_i p_i^2$ with $b_1= 2B,b_2=2A$, see (\ref{48c}).
We simplify below this expression, writing $\exp(-b_i p^2)= \frac{1}{\exp(b_i p^2)}\approx b_i^{-1} \frac{1}{p^2 +\nu_i^2}$, where $\nu_i=\frac{1}{\sqrt{b_i}}$.
As a result one can write for $I_i(E)$ in the region $E > E_i(th)=m_1+m_2 (i=1)$ or $m_3+m_4 (i=2)$
\be
I_i(E)= \frac{\mu_i}{2\pi b_i }\nu_i -i \sqrt{2\mu_i\Delta_i}, \Delta_i= E - E_i(th).
\label{6.6} \ee
As a result one obtains a simple expression for the amplitude $f_{12}(E)$

\be f_{12}(E)= \frac{1}{1- \frac{\lambda' \mu_1\mu_2}{(\nu_1 - i\sqrt{2\mu_1\Delta_1})(\nu_2 -i\sqrt{2\mu_2\Delta_2)}}}
=\frac{1}{1-z t_1(E) t_2(E)}.
\label{6.7}\ee
Here $z= \frac{\lambda}{\mu_1 \mu_2 b_1b_2 (2\pi)^2}$ and the (\ref{6.7}) refers to the region $E > E_1({\rm th}),E_2({\rm th})$
otherwise one should replace $-i$ before the $\sqrt{}$ terms by $(+1)$ for the roots, where $\Delta(E)$ is negative.

We now consider the channel coupling constant $\lambda'$, which enters (\ref{6.7}).  From (\ref{48c}) one obtains
\be \lambda'= \frac{ M_\omega^2 \chi^2_{1234} b_1 b_2}{4 \pi^5 a^3 (\prod_i \beta_i)^3}.
\label{6.8} \ee
The resulting $z$ is larger than unity for the recoupling coefficient $\chi_{1234}$ of the order of 1, and
one can vary $z$ in the interval from one to larger values.
At this point we shall discuss the physical structure of the obtained expression for the amplitude $f_{12}(E)$ (\ref{6.7}). One can see that $f_{12}$ contains the product of two unitary single channel amplitudes $t_1,t_2$
which describe separately scattering amplitudes in channel 1 and 2 characterizing by numbers $\nu_i$, $i=1,2$.
The latter are obtained from the hadron wave functions and reflect their structure given by the Gaussian coefficients $\beta_i$. Now let us study the structure of the possible singularities of $f_{12}$ in the $E$ plane, which is given
by zeros of the denominator of $f_{12}$ in (\ref{6.7}) and we concentrate on the $k_1$ plane while the second channel is characterized by the value of $B(E)= \frac{1}{\nu_2 -ik_2}$ and $-ik_2 >0 $ below higher threshold $E_2$. As a result
one obtains an equation for the pole of $f_{12}$
\be
 k_1= -i\nu_1 +i z B(E) , B(E) >0, \nu_1> 0.
\label{6.9}\ee
It is evident that originally (at $zB=0$) virtual pole of $k_1$ is proceeding
 to the upper half plane of $k_1$ with increasing $zB$, and at some $zB$ at appears at $k_1=0$ i.e. at the threshold $E_1$, and at larger $zB$ one obtains a real pole, i.e. the effect of recoupling with the channel 2 increases
 the attraction and creates a real pole from the original distant virtual pole.
However for the physical amplitude $t_1(E)$ of the hadron-hadron interaction  the virtual pole can be not a good
approximation and more generally one should write for the amplitude $t_1(E)=\frac{1}{\nu_1(E) -i k_1}$ the $\nu_1(E)$ in the form $\nu_1(E)= a + b ln(\frac{m^2+4 k_1^2}{m^2})$, where $a,b$ are positive numbers. This construction
occurs from the t-channel exchange of the hadron with mass $m$. One can see from (\ref{6.9}) that the logarithmic
singularity is moving towards the threshold with the increasing $z$ and can appear on the first sheet ($E<E_1$).
We can now study the situation with the singularities associated with the threshold $E_2$ and find that here the situation is completely different because when $E~E_2$ the momentum $k_1(E)$ as a rule is large and positive imaginary. As a
result there is no pole solutions of the equation (\ref{6.7}) for all values of $E$ nearby the threshold $E_2$ and
all resulting  widths are of the order of parameters $\mu_i,\nu_i$, i.e. $(0.5-1)$ GeV, and therefore cannot be seen
in experiment. Nevertheless we give in the Apendix 4 full quartic equation and its analysis for the finding of resonance poles in the whole combination of $k_1,k_2$ planes. As a special case one can consider the situation when
the distance $\Delta=E_2-E_1$ is small as compared with $\nu_i,i=1,2$. In this case the amplitude $f_{12}(E)$ written as
\be
f_{12}= \frac{k_2-i\nu_2}{k_2 +\frac{zk_1}{\nu_1^2+k_1^2} +i(\nu_2 -\frac{z\nu_1}{\nu_1^2+k_1^2})}.
  \label{6.10}
\ee

  One can see that for $k_1<<\nu_1$ and $\nu_2~\frac{z}{\nu_1}$ one obtains a resonance pole slightly shifted above the threshold $E_2$, however this situation is coincidental. Summarizing we expect that the resulting singularities  of the recoupling amplitude are produced by the singularities of the lower threshold process (process 1 in our definition) shifted to the lower threshold due to recoupling interaction with the process 2). In the next section we shall study this effect numerically in real physical examples.

\section{Numerical results and discussion}
 We can now consider 3 transitions, partly discussed above:

 \begin{description}

\item{{\bf 1)}}  $$ J/\psi + \phi \to D^*_s + \bar D^*_s $$
 $E_1({\rm th})= 4.12~ {\rm GeV},~ E_2({\rm th})= 4.224~{\rm GeV}, \mu_1= 0.767, \mu_2= 1.056, \nu_1= 0.96, \nu_2= 0.87, z=1$ (all GeV).

 \item{\bf 2)} $$ J/\psi + \phi \to D_s + \bar D_s $$
 $E_1({\rm th})= 3.936, E_2({\rm th})= 4.12, \mu_1= 0.767, \mu_2= 0.984, \nu_1= 0.96, \nu_2= 0.87, z= 1$.

  As a special interesting case we consider below the recent experiment of BES III \cite{66},
  $$ e^{+} + e^{-} \to K^{+} ( D_s^{-} D^* + D^{*-}_s D). $$

Applying here our recoupling mechanism, shown in  the Fig. 1, one easily finds that the second channel obtained from the first channel
$D_s^{-}D^*$ by recoupling is the channel $J/\psi + K^{*-}$ which creates the chain of reactions possibly
  generating a peak in the system $D_s^- D^*$ or $D^{*-}_s D$, namely we consider as the third example

 \item{\bf 3)} $$ D_s^{-} + D^*_s \to J/\psi + K^{*-} $$
 $E_1({\rm th})= 3.975, E_2({\rm th})= 3.992, \mu_1= 0.9936, \mu_2= 0.692, \nu_1= 0.87, \nu_2= 0.96$ (all GeV).

 \end{description}

 One can easily find that in all cases the values of $\nu_i$ and the values of $\mu_i$ are in the
 range $0.35-1.06$ GeV. Applying the (\ref{6.7}) one can find the $f_{12}$ in all cases and hence $\frac{d\sigma}{d E}$
 in the cases {\bf 1), 2), 3),}  assuming $z$ as a positive number typically larger or equal 1. However for the {\bf 3)} one could use also the sum:
 $f_{12} \to f_{12} + \alpha f'_{12}$, implying possible superposition of intermediate states in the rescattering series.

We proceed now with the cases {\bf 1)-3)} and insert the values of $E_i(th),\mu_i,\nu_i$ in (\ref{6.3},\ref{6.7}) and fixing the value of $z$ one obtains the form of the recoupling amplitude shown in the Tables II-IV below.

For the case  {\bf  1)}  the resulting values of $|f_{12}(E)|^2$ can be seen in the Table 2.

\begin{table}[h!]
\caption{The values of the $|f_{12}(E)|^2$ near the channel thresholds for the transition {\bf 1)} }
\begin{center}
\label{tab.02} \begin{tabular} {|c|c|c|c|c|c|c|} \hline
$E$(GeV)& 4.0& 4.05& 4.12& 4.17& 4.224& 4.3\\
$|f_{12}(E)|^2$& 3.43&4.49& 17.36& 10.76& 5.6& 1.24\\\hline
\end{tabular}
\end{center}
\end{table}

One can see in the Table II a strong enhancement around $E= 4.12$ GeV with the width around $10$ MeV which can be associated with the resonance
$\chi_{c1}(4140)$ having the mass $4147 $ MeV and the width $ \Gamma= 22  $ MeV.

It is now interesting how these numerical data are explained by the exact solution of the equation (\ref{6.7},\ref{6.9}).
From these equations one obtains the exact position of the pole $k_1=i 0.068$ GeV, $E= E_1- 0.003$ GeV which is unphysical result due to too small values of $\nu_1$, and in a more realistic case of the virtual pole its position moves to $k_1=0,E=E_1$ for $\nu_1= 0.84$ GeV. Three more complex poles are O(1 GeV) far from the thresholds, as can be found from the solution of the quartic equation in Appendix 4. Looking at table II one can see a good agreement with this
result.
In a similar way we obtain the results for the $J/\psi + p$ transitions of {\bf 2)}.

\begin{table}[h!]
\caption{The values of the $|f_{12}(E)|^2$ near the the $\Sigma D$ threshold}
\begin{center}
\label{tab.03}
\begin{tabular}{|c|c|c|c|c|c|c|c|} \hline
$E$(GeV)& 3.8& 3.85& 3.936&4.0& 4.05& 4.12&4.2\\
$|f_{12}(E)|^2$&3.12& 3.84& 18.9& 6.15& 5.33&298&1.17\\\hline
\end{tabular}
\end{center}
\end{table}

One can see in Table III a strong peak near the lower threshold $E=3.94$ GeV and it is easy to check that this peak is stable when one varies $z$ in the region around $z=1$, One can associate this peak with the virtual pole
appearing at 50 MeV below the threshold $E_1= 3.94$ GeV. The exact form of the solution for $f_{12}(E)$ near the peak has the form
$f_{12}(E)= \frac{k_1 + i0.96}{k_1 + i0.28}$.   In this way our method
can support the origin of the $X(3915)$ state as due to the $J/\psi + \phi \leftrightarrow D_s + \bar D_s$ transitions.

We come now to the recent interesting discovery of the new state $Z_{cs}(3985)$ \cite{66}, where we take for simplicity only the first chain denoted as the {\bf 3)} above. Similarly to the previous cases one obtains

\begin{table}[h!]
\caption{The values of the transition probability as a function of energy in the transition {\bf 3)} }
\begin{center}
\label{tab.04} \begin{tabular} {|c|c|c|c|c|c|c|} \hline
$E$(GeV)& 3.96& 3.975& 3.98& 3.985& 3.992& 4.0\\
$|f_{12}|^2(z=1)$& 5.43& 322& 67.56&31.84& 12.95& 5.54\\
$|f_{12}|^2(z=1.5)$& 19.9& 2.93& 3.19& 2.58& 1.61& 2.71\\\hline
\end{tabular}
\end{center}
\end{table}


One can see in Table IV a narrow peak with the summit at $E=3.975$ GeV for $z=1$ with the width around $10$ MeV,
which closely corresponds to the experimental data from \cite{66} $E=3.982.5,\Gamma= 12.8$ MeV.
In our case the resonance parameters weakly depend on $z$. As seen in the Table IV, for $z=1.5$ the peak shifts
down to $E= 3.960$ GeV with much larger width. Its form is exactly the same as in the previous two examples and corresponds to virtual pole almost exactly at the lower threshold. In this way we can explain the newly discovered resonance $Z_{cs}(3985)$ by the recoupling mechanism in the
rescattering series of transitions $$ D_s^{-} + D^*_s \to J/\psi + K^{*-} $$.

\section {Conclusions and an outlook}

As it was shown above, the new mechanism having the only parameter $z$ is able to predict and explain the resonances
in different systems, as it was shown above, and possibly in other systems which can transfer
one into another via the recoupling of the confining strings. The necessary conditions for the realization of these
transitions and the appearance of a resonance are connected to the value of the transition coefficient $z$, which should be of the order of unity or larger. Therefore the transition should be strong, i.e. without serious restructuring of the hadrons involved, since otherwise the transition will be strongly suppressed e.g. in the case
when not only strings are recoupled, but also spins,orbital momenta, isospins should be exchanged.
In any case the suggested mechanism provides an alternative to the popular tetra -- and pentaquark mechanisms, which
dominate in the literature. One should stress at this point, that the independent and objective checks, e.g. the
lattice calculations do not give strong support for the molecular or tetraquark models and the existence of an independent mechanism is welcome. It is necessary that this method  should be
studied more carefully.
As to the recoupling mechanism, it is strongly associated with the thresholds participating in the transitions,
and the best situation for its application is when both thresholds are close by. In all 3 cases considered above the distance between threshold was less than 200 MeV, and in all cases one could see a strong enhancement in the transition coefficient and hence in the resulting cross section.
The necessary improvements of the present study are 1) a more accurate calculation of the coefficient $z$ (originally $\lambda'$), and 2) the use of a more realistic Gaussian approximation for the wave functions instead of approximate     $\nu_i$ parametrizations  to define $f_{12}$ with good accuracy in the future.
The author is grateful to Lu Meng for an important remark, to R.A.Abramchuk for useful discussians and to A.M.Badalian for discussions and suggestions.
 This work was done in the frame of the scientific project supported  by the Russian Science Foundation Grant No. 16-12-10414.

  \vspace{2cm}

{\bf Appendix A1.}  {\bf The  center-of-mass coordinates and average quark and antiquark energies in a hadron}  \\

 \setcounter{equation}{0} \def\theequation{A1.\arabic{equation}}

Following \cite{30**} one can define the c.m. coordinate of a  hadron  consisting  of a quark $Q$ at the point $\vex$ and an antiquark $\bar q$ at the point $\veu$ via average energies $\Omega$ and $\omega$  of$Q$ and $\bar q$ correspondingly as
\be \veR_{Q\bar q} = \frac{\Omega \vex + \omega \veu}{\Omega + \omega}, ~~  \Omega = \lan  \sqrt{ \vep^2_Q + m^2_Q}\ran, ~~ \omega = \lan \sqrt{ \vep^2_q + m^2_q}\ran\label{A1.1}\ee where $\sqrt{ \vep^2_Q + m^2_Q} + \sqrt{ \vep^2_q + m^2_q} $ is the kinetic part of the $Q\bar q$  Hamiltonian in the so-called spinless Salpeter formalism or an  equivalent form in the so-called einbein formalism.\footnote{  see a short recent review in  the last refs. in \cite{30*}}

As a result one obtains the following value of $\omega=\Omega$ for $q\bar q$ mesons,  shown in Table II.

 \begin{table}[!h]
\caption{Average values of quark and antiquark kinetic energies in different mesons }
 \begin{center}
\label{tab.05}\begin{tabular}{|c|c|c|c|c|c|}\hline

State & $J/\psi$ & $\psi (2S)$ & $\psi (3770)$ & $\psi(3S)$ &$\psi (4S)$\\ \hline

$\Omega $, GeV& 1.58 &1.647 & 1.640 & 1.711 & 1.17 \\\hline

State & $\Upsilon (1S)$ & $\Upsilon (2S)$ & $\Upsilon (3S)$ & $\Upsilon (4S)$ &$\Upsilon (5S)$\\ \hline

$\Omega $, GeV& 5.021 &5.026 & 5.056 & 5.088 & 5.120 \\\hline

Sate& $D$ & $D_s$&$B$&$B_s$& $\rho$    \\\hline

$\Omega $, GeV& 1.509 &1.515 & 4.827 & 4.830  &0.4  \\
 \hline

$\omega $, GeV& 0.507 &0.559& 0.587 & 0.639 &0.4  \\
 \hline
\end{tabular}

\end{center}*) Note, that the difference in $\Omega, \omega $ obtained in these approaches is less or around 1\%.
\end{table}

  \vspace{2cm}

{\bf Appendix A2.}  {\bf The channel-coupling coefficient $\bar y_{1234}$}  \\

 \setcounter{equation}{0} \def\theequation{A2.\arabic{equation}}

We discuss here two topics: 1) the problem of the double string decay vertex  contribution to the recoupling  coefficient $M_{\omega}$  in (\ref{40}), 2)  the   construction of the  recoupling vertex $\bar y_{1234}$.

We start with the topic 1), and following  \cite{33***}  define the relativistic expression for  the string decay vertex as in (\ref{38}), but without free parameters  namely $M(x)$ in (\ref{38}) is
\be  M(x) =\sigma (|\vex -\vex_{Q }|+ |\vex-\vex_{\bar Q}|).\label{48}\ee

As one can see in Fig.5, in our case the structure of the  recoupling process can be  explained by the double string breaking, which we can write as a product
\be S=\int d^4x \bar \psi (x) \bar M (x)  d^4y  \psi (x) \bar \psi (y) \bar M (y) \psi(y)\label{49}\ee
and one must take into account,  that the energy minimum of the  resulting broken string occurs when both time moments $x_0,y_0$  of string breaking are equal. Indeed, taking the integral in (\ref{49})   with account of the string action in the  exponent of the path integral,$$ \Delta S_{\rm string} = \int \sigma \sqrt{ r^2_{xy} + (x_4-y_4)^2}  d\frac{x_4+y_4}{2},$$ which produces a factor on (\ref{49}) $\lan r_{xy}\ran \sqrt{2\pi}$, which  denoted  as  $M_\omega$ in (\ref{40}).  The resulting double string breaking action can be written as,

\be S\cong \int d^3 x \bar psi (x) \psi (x) d^3y \bar \psi (y) \psi (y)   M_\omega (\vex,\vey) \frac{d(x_4+y_4)}{2}\label{50}\ee
using the notation  $$\int d (x_4 - y_4) \lan \bar M(x) \bar M(y)\ran = M_{\omega} (\vex, \vey).$$

In what follows one can  estimate $\lan M_\omega (\vex,\vey)\ran \equiv M_\omega$ in the same way, as it was done in \cite{33***}, with the result $M_\omega \approx \frac{2\sigma}{\beta}$, where $\beta$ is the oscillator parameter for the $(Q\bar Q)$ meson. We now turn to the point 2) above, the recoupling vertex $\bar y_{1234}$.

To define $\bar y_{1234}$  we  notice that  all 4 quarks  $q, \bar
q, Q,\bar Q$ keep their identity and spin polarization during the whole process of transformations, provided  we neglect the  spin dependent corrections. This can be also seen in the structure of the recoupling process: in (\ref{49}) one does  not   see spin  dependence, and this means, that the spin projection of each quark or antiquark is kept unchanged during recoupling. As a result one can write the nonrelativistic spin part of the matrix element $V_{12/34}$ as
\be V_{meson}^{spin} = C^{Lm}_{\mu_1\mu_2} \chi^{(1)}_{\mu_1} \bar \chi_{\mu_2}^{(2)}
C^{L'm'}_{\mu'_1\mu'_2}
\chi^{(3)}_{\mu'_1} \bar \chi_{\mu'_2}^{(4)}
(C^{JM}_{\nu_1\nu_2} \chi^{(5)}_{\nu_1} \bar \chi_{\nu_2}^{(6)}
C^{J'M'}_{\nu'_1\nu'_2}
\chi^{(7)}_{\nu'_1} \bar \chi_{\nu'_2}^{(8)})
\label{51}\ee
where the Klebsch-Gordon coefficient $C^{JM}_{\nu_1\nu_2} \equiv C_{\frac12\nu_1\frac12\nu_2}$ and $\chi^{(i)}_\mu, \bar \chi^{(k)}_\lambda$ are quark and antiquark spinors.

As was told above, due to the spin  conservation in   recoupling, the matrix element (\ref{51}) should be proportional to $\delta_{26}\delta_{48}\delta_{17} \delta_{35}$,  implying the  the recoupling of quarks.

As a result one obtains
\be V^{spin} =\sum_{\mu_i,\mu'_k} C^{Lm}_{\mu_1\mu_2}  C^{L'm'}_{\mu'_1\mu'_2}  C^{JM}_{\mu'_1\mu_2}  C^{J'M'}_{\mu_1\mu'_2}.\label{52}\ee
 Here $L,L',J,J'$ correspond to the spin values of hadrons $L+L'\to J+J'$ and we have assumed zero orbital momenta for all hadrons. Finally for the final expression in (\ref{40}), $(V^{\rm spin})^2$ should be summed up over all $ m,m' M,M'$, so that for $\bar y^2_{1234}$ one has \be \lan (\bar y_{1234})^2\ran = \sum_{m,m'M,M'}(V^{\rm spin})^2.\label{53}\ee

 In a similar way one can find the recoupling coefficient $\bar y_{1234}$ for the  ensemble transformation
 \be p+J/\psi\to \Sigma  (\Sigma^*) + D(D^*) \to p + J/\psi.\label{54}\ee

 In this case we write the  baryon  wave function as  $\psi_B=   u (ud)_0  $, where the lower indices imply the total spin of the diquark ($ud)$.
 In the simplest approximation one can approximate the  proton as the quark-diquark combination $p=u(ud) \cong u\tilde{d}$, with the diquark $\tilde d$ kept unchanged $d$ during  recoupling.

  \vspace{2cm}

{\bf Appendix A3.}  {\bf Oscillator parameters of hadron wave function}  \\

 \setcounter{equation}{0} \def\theequation{A3.\arabic{equation}}

 The oscillator parameters for the bottomonium, charmonium and $B,D$ mesons have been obtained in \cite{23*,30*}, using the expansion of  relativistic wave  functions, obtained from the solutions of  the  relativistic string Hamiltonian \cite{30**}, in the full set of the oscillator wave functions. As a result one obtains

 \begin{table}[!h]
\caption{The Gaussian parameters $\beta$ of different mesons }
 \begin{center}
\label{tab.06}\begin{tabular}{|c|c|c|c|c|c|}\hline

State & $\Upsilon (1S)$ & $\Upsilon (2S)$ & $\Upsilon (3S)$ & $\Upsilon (4S)$ &$\Upsilon (5S)$\\ \hline

$\beta $, GeV& 1.27 &0.88 & 0.76 & 0.64 & 0.6 \\\hline

State & $J/\psi$ & $\psi (2S)$ & $\psi (3S)$ & $\psi(4S)$ &$\psi (5S)$\\ \hline

$\beta $, GeV& 0.7 &0.53 & 0.48 & 0.43 & 0.41 \\\hline

Sate& $D$ & $B$& $\rho$ &   &   \\\hline

$\beta $, GeV& 0.48 &0.49 & 0.26 &   &  \\
 \hline
\end{tabular}

\end{center}
\end{table}

The accuracy of the oscillator one-term approximation can be judged by the relative value   of the sum od squared coefficients of four higher term of expansion as compared to the square of the main term. This amounts to the accuracy of the order or less than 10\% for lowest states of charmonia and bottomonia and few percent for $D,B,\rho$.

\vspace{1cm}
{\bf Appendix A4} {\bf Calculation of the cross sections $|f_{ik}|^2$ and pole positions }\\

 \setcounter{equation}{0} \def\theequation{A4.\arabic{equation}}

As it is written in (\ref{6.7}) the cross section is defined as $\frac{d\sigma}{dE} = |f_{ik}|^2$ where
$f_{ik}= \frac{1}{1- z A B}$, and z is numerical parameter proportional to $\lambda'$,  where for three
numerical examples $(1),(2),(3)$ in that section $z= 1, 5, 1$ and $A=\frac{1}{\nu_1 - i k_1}$ and
$B= \frac{1}{\nu_2 - i k_2}$. Moreover $ k_i= \sqrt{2\mu_i(E- E_i)}$ and parameters $\mu_i,\nu_i,E_i$  are
given in the section. As it is seen in Tables II,III,IV the cross sections have peaks and our purpose here is to
calculate the positions of the poles both in E plane and in $k_i$ plane. To do this one starts with two equations
\be
 (\nu_1 -i k_1)(\nu_2 -i k_2)=z, \frac{k_1^2}{2\mu_1} - \frac{k_2^2}{2\mu_2}= E_2 - E_1.
 \label{56}\ee

 One of possible strategy is to insert $k_1$ from the first equation into second equation and one obtains an equation of 4-th power for $k_2$. Note that both $k_i$ are positive on the real axis for $E > E_i$ and positive imaginary below $E_i$. Among many roots of eq. we need the closest to the real axis.
 The pole positions are below in the Table.
Now defining from the first equation in (\ref{56}) $k_1$ via $k_2$ and submitting it into second equation one
obtains the quartic equation for $x= k_2 + i\nu_2$
\be
x^4 + a x^3 + b x^2 + c x + d =0
\label{57}, \ee
where the coefficients are
\be
a= -2i\nu_2, b= 2\mu_2 \Delta - \nu_2^2 + \frac{\mu_2 \nu_1^2}{\mu_1}, c= -2i\frac{\nu_1\mu_2 z}{\mu_1},
d= - \frac{mu_2 z^2}{\mu_1}.
\label{58} \ee
Now taking values of these coefficients from item{\bf i)}, for three transitions $ i= 1,2,3$ discussed in section {Numerical examples and discussion} with $z= 1, 0.2 ,1$ respectively for these transitions, and $\Delta=
E_2 - E_1$, one obtains the roots of this equation  together with the corresponding values of the
resonance energy. The results of this computation for the examples 1)-3) are given in the text of the section 7.

\end{document}